\documentclass[12pt,preprint]{aastex}

\shorttitle{X-ray Spectroscopy of Vela Shrapnel A}
\shortauthors{Katsuda et al.}

\begin{document}

\title{Spatially Resolved X-ray Spectroscopy of Vela Shrapnel A}

\author{S. Katsuda\altaffilmark{1} and H. Tsunemi\altaffilmark{1}}

\altaffiltext{1}{Department of Earth and Space Science,
	Graduate School of Science, Osaka University,\\
	1-1 Machikaneyama, Toyonaka, 560-0043 Osaka, Japan; 
	katsuda@ess.sci.osaka-u.ac.jp, tsunemi@ess.sci.osaka-u.ac.jp,}

\begin{abstract}

  We present the detailed X-ray spectroscopy of Vela shrapnel A with the XMM-Newton satellite.  Vela shrapnel A is one of several protrusions identified as bullets from Vela supernova explosion.  The XMM-Newton image shows that shrapnel A consists of a bright knot and a faint trailing wake.  We extracted spectra from various regions, finding a prominent Si Ly$_\alpha$ emission line in all the spectra.  All the spectra are well represented by the non-equilibrium ionization (NEI) model.  The abundances are estimated to be O$\sim$0.3, Ne$\sim$0.9, Mg$\sim$0.8, Si$\sim$3, Fe$\sim$0.8 times their solar values.  The non-solar abundance ratio between O and Si indicates that shrapnel A originates from a deep layer of a progenitor star.  We found that the relative abundances between heavy elements are almost uniform in shrapnel A, which suggests that the ejecta from supernova explosion are well mixed with swept-up interstellar medium. 

\end{abstract}


\section{Introduction}
  The nucleosynthesis process inside a star generates high-Z elements, and supernova explosions spew them out into interstellar space.  Fragments of the ejecta have been found in some supernova remnants (SNRs) e.g., fast-moving knots in Cas A (e.g., Kirshner \& Chevalier 1977; Fesen et al. 2002), southeastern knots in Tycho SNR (Decourchelle et al. 2001) and several X-ray emitting regions in the vicinity of the Vela SNR (Aschenbach et al. 1995, hereafter AET).
  
  The Vela SNR is one of the most famous SNR in the X-ray sky. The age is estimated to be $\sim$11400 years old (Taylor et al. 1993) and the distance to the SNR is estimated to be $\sim$250 pc (Cha et al. 1999).  It is so close to us that the angular diameter of the Vela SNR is very large, making it an ideal target for studying fine structures of the SNR.  The ROSAT all sky survey revealed that the Vela SNR had an almost circular appearance with a diameter of 8.$^\circ$3 (AET).  There are a significant amount of protruding features beyond the primary blastwave.  Some of them have boomerang structures whose opening angles suggest supersonic motion in a tenuous matter.  The center lines of the boomerang structures intersect near the center of the Vela SNR.  Therefore, AET concluded that those features are bow-shocks created by bullets which overran the blastwave.  They named those features 'shrapnels A $\sim$ F'.  If they are really debris of the SN ejecta, the spectra of shrapnels would show a metal-rich composition.  So far, the X-ray satellites ASCA (Tsunemi et al., 1999, hereafter TMA), Chandra (Miyata et al., 2001, hereafter MTAM) and XMM-Newton (Katsuda \& Tsunemi, 2005) observed some of them and revealed their nature.  Furthermore, Chandra or XMM-Newton enables us to perform spatially resolved analysis for those shrapnels. 

  Shrapnels A and D are the best investigated targets among the seven shrapnels.  Recently, XMM-Newton observed shrapnel D which is the brightest and largest shrapnel of them.  Katsuda \& Tsunemi (2005) found extreme over abundances of O, Ne and Mg while the abundance of Fe is about solar value; the relative abundance to the solar values are O$\sim$5, Ne$\sim$10, Mg$\sim$10.  Since the Vela SN explosion is believed to be a type {\scshape II} explosion, they concluded that shrapnel D is an explosion ejecta from the Vela SN explosion (probably from an O, Ne and Mg rich-layer).  Inside the field of view (FOV) of XMM-Newton, the abundances of heavy elements were uniform, although it has so large of an angler size that even XMM-Newton with a large (the diameter of 30$'$) FOV could not cover the entire X-ray emitting region of shrapnel D.  TMA observed shrapnel A with ASCA.  They found a prominent Si K emission line with relatively weak emission lines from other elements, revealing that the relative abundance of Si is a few times ten higher than those of other elements.  Therefore, they concluded that shrapnel A was an explosion ejecta from a Si-rich layer of a progenitor star.  The spatial resolving power of ASCA was not enough for them to perform spatially resolved analysis.  In order to perform spatial resolved analysis, Miyata et al. (2001) observed it with Chandra.  The Chandra image reveals a bright X-ray region at the head of the shrapnel and a fainter extended tail.  They confirmed an overabundance of Si in the head, but could not obtain a strong constraint on the parameters of the tail.  

  We analyzed archival data of shrapnel A observed with XMM-Newton.  Owing to the good spatial resolving power and the large effective area of XMM-Newton, we are able to perform spatially resolved spectral analysis of the shrapnel.

\section{Observations}

  XMM-Newton observed shrapnel A on 2000 December 7, 8 with the EPIC (European Photon Imaging Camera) instruments (Obs ID is 0112870101).  The EPIC MOS cameras were operated in the standard full-frame mode and EPIC PN camera was operated in the extended full-frame mode.  The EPIC MOS-1 and PN camera were operated using a thin filter while EPIC MOS-2 camera was operated using a medium filter.  Since there were high-background periods for MOS-1 and PN camera, we screened the data by rejecting the periods.  As a result, the effective exposure time was reduced to $\sim$ 45 ks for MOS-1, to $\sim$ 48\,ks for MOS-2 and $\sim$ 30 ks for PN.  We selected X-ray photons corresponding to patterns 0--12 for MOS and patterns 0 for PN, respectively.  In order to subtract background data from the same region as the source region on the CCD, we used observations of the Lockman hole as background data.  We used observation 0123700401 for MOS-1 and PN and observation 0147511701 for MOS-2, respectively.  Note that we had to select background data such that the optical blocking filters used correspond to those used in the observation of shrapnel A.

\begin{figure}
\includegraphics[angle=0,scale=.4]{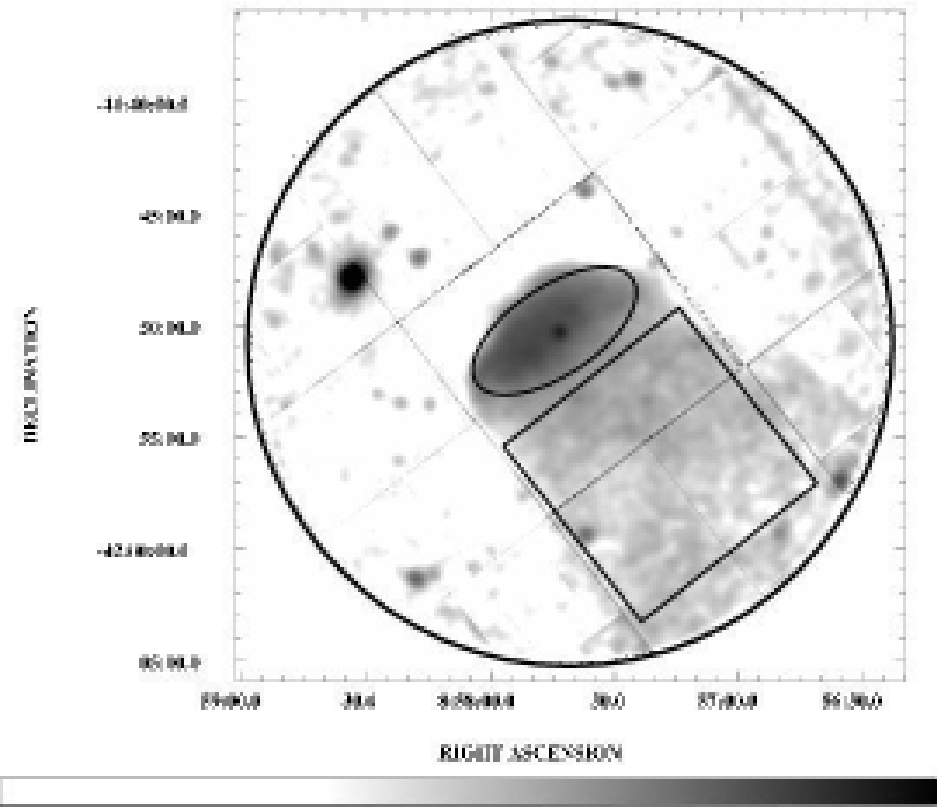}
\includegraphics[angle=0,scale=.4]{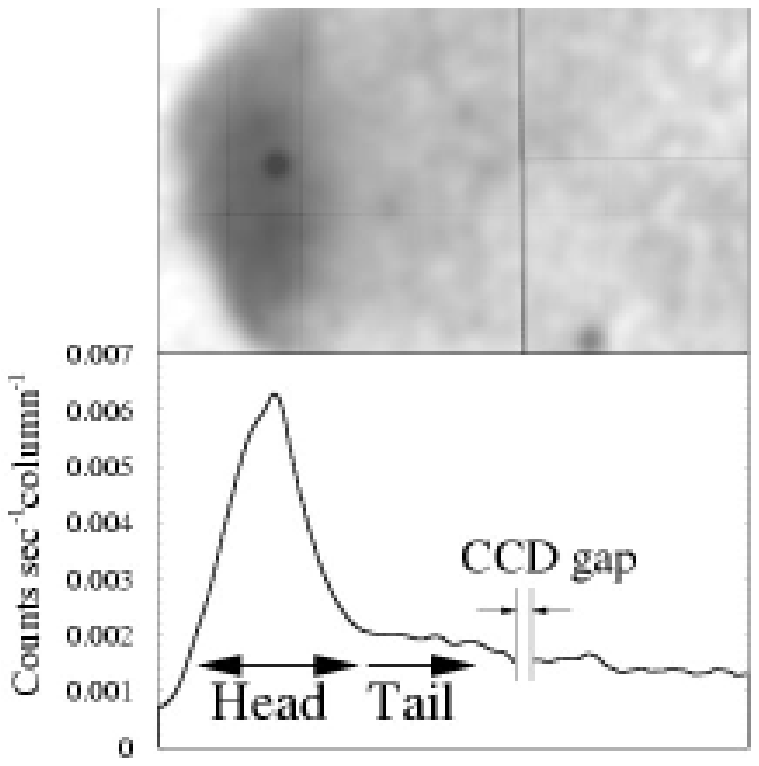}
\caption{Left: EPIC MOS-1\,$+$\,MOS-2 logarithmically scaled image in the energy range 0.2--3.0\,keV.  The data have been smoothed with a Gaussian of $\sigma = 12''$.  The spectra from the ellipse and square are shown in figures \ref{spec_head} and \ref{spec_tail}.  Right (upper): Same as figure \ref{image1} left but focused on shrapnel A and rotated such that the moving direction becomes horizontal.  Right (lower): The X-ray intensity projected from the upper image.  Note that obvious point sources were not removed from the profile.}
\label{image1} 
\end{figure}

  Figure \ref{image1} left shows an exposure-corrected MOS-1\,$+$\,MOS-2 image of shrapnel A that is seen as a triangular-shaped feature located from the center to the southwestern edge of the FOV.  We extracted the image in the energy range of 0.2--3\,keV and extracted the exposure map using SAS v\,5.4.1(the EEXPMAP command).  We smoothed the image by a Gaussian of $\sigma$ = $12^{\prime \prime}$.  We can see a bright X-ray knot feature and extended trailing emission toward the main shell of the Vela SNR as MTAM found by Chandra.  Figure \ref{image1} right (top) is the image of shrapnel A along the direction to the center of the Vela SNR.  Figure \ref{image1} right (bottom) shows the projection of the intensity of the upper image.  We can see a bright knot region as well as a trailing part.  Hereafter, we define the bright region as the head and the following region as the tail as shown in figure \ref{image1} right (bottom).  There is also several point sources in the FOV.  The brightest one is CU Velorum, located at the eastern edge of the FOV.  

\section{Spectral Analysis}
\subsection{CU Velorum}

  CU Velorum is classified as a non-magnetic cataclysmic variable (Baskill et al. 2005).  The spectrum obtained by ASCA is well represented by a single-temperature thin thermal plasma model (the MEKAL model, Mewe, Lemen \& van den Oord 1986; Liedahl, Osterheld \& Goldstein 1995).  The temperature determined from the spectral fit is 3.5$^{+0.6}_{-0.5}$\,keV.  We confirmed that our spectrum is also well represented by the same thin thermal plasma model.  We also confirmed that the electron temperature obtained to be 4.24$^{+0.17}_{-0.20}$\,keV is consistent with that obtained by ASCA.  The flux obtained in the energy range from 0.8\,keV to 10\,keV is 1.84$\times$10$^{-12}\mathrm{ergs\,cm^{-2}\,sec^{-1}}$ which is about 1.7 times higher than that obtained by ASCA.  Using the distance of 209 pc (Warner B. 1987, Gansicke et al. 1999), the luminosity (0.2 -- 10.0\,keV) is estimated to be 1.2$\times$10$^{31}$ erg\,sec$^{-1}$.  Since non-magnetic cataclysmic variables tend to have luminosities of 10$^{30}$--10$^{32}$ erg\,sec$^{-1}$ (Verbunt et al. 1997), the observed luminosity is a typical value.     

\subsection{Shrapnel A}

  We investigated spectral variations from various regions in detail.  We extracted spectra from a rectangle of 2$^\prime\times$2$^\prime$ for the head region and from that of $4^\prime\times10^\prime$ for the tail region, respectively.  The size was selected such that we could obtain sufficient number of photons in each rectangle (1700$\sim$4500 photons for the MOS-1 detector).  We arranged rectangles such that each rectangle overlapped each other by half of its size and all the rectangle fully cover shrapnel A, obtaining 24 spectra for head region and 5 spectra for tail region in total.  We removed visible point-like sources for the spectral analysis.  We have used the Science Analysis System (SAS, version 6.1.0, version 5.4.1 only for the exposure map) for data reduction. 

\subsubsection{Head Region}

\begin{figure}
\includegraphics[angle=0,scale=.3]{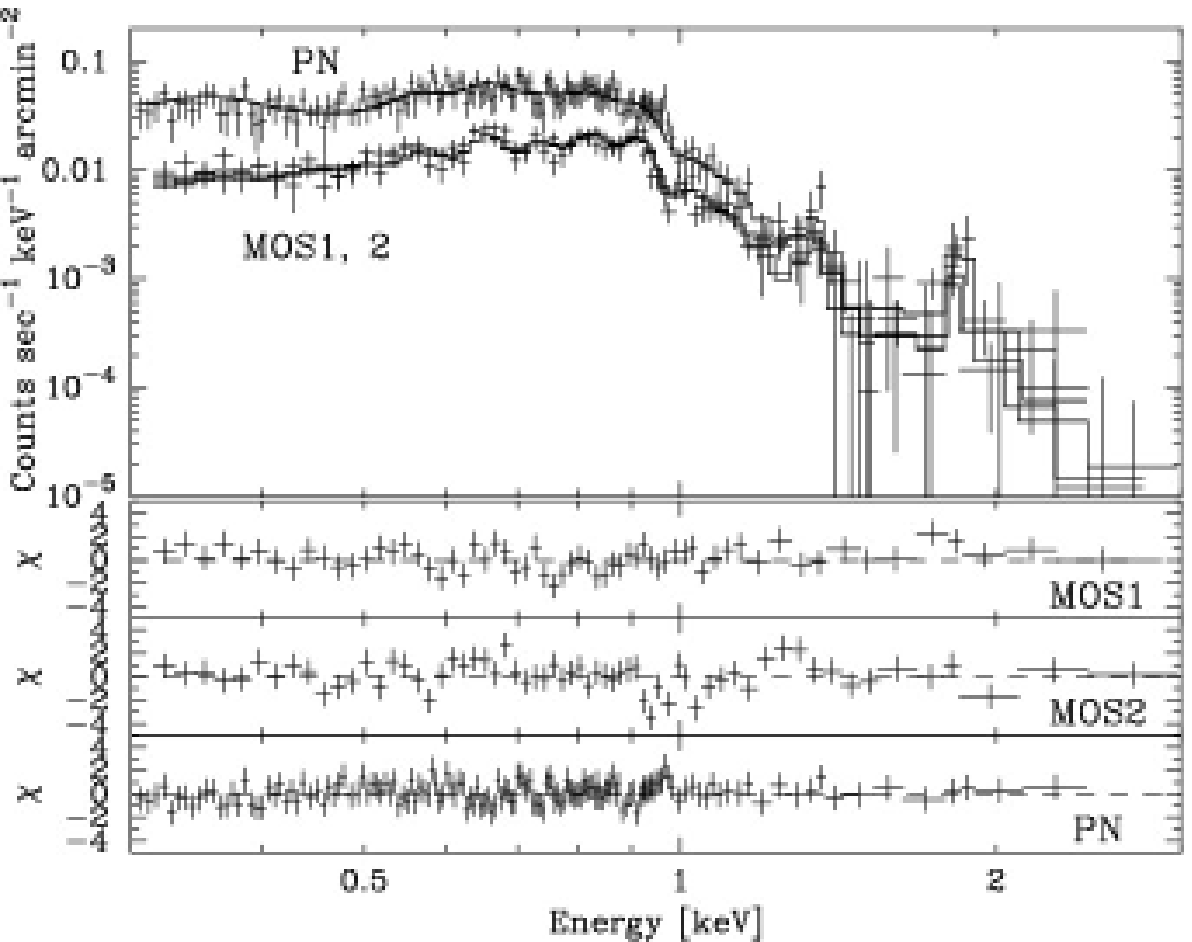}
\includegraphics[angle=0,scale=.3]{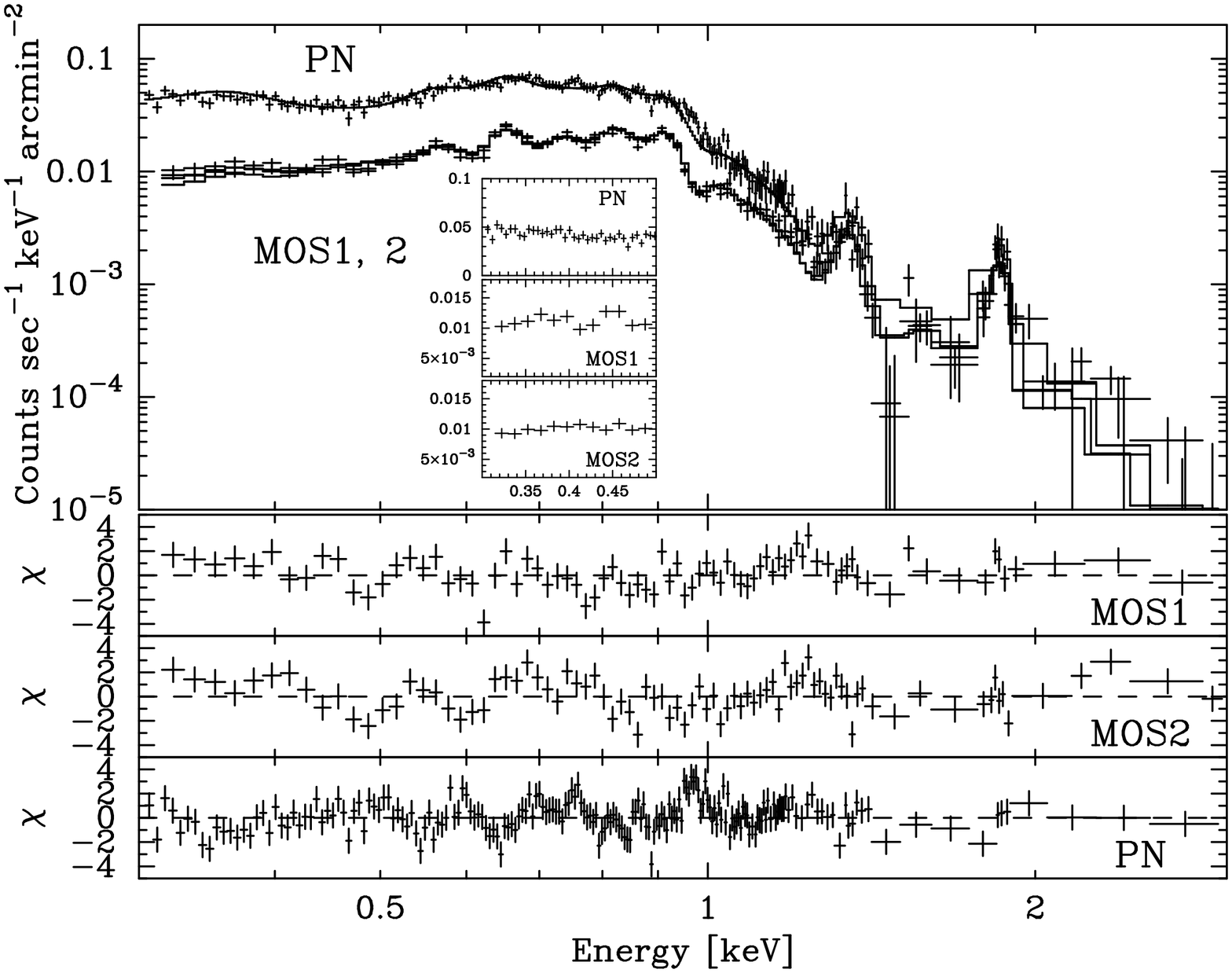}
\caption{X-ray spectra extracted from the head region.  Left panel shows an example spectrum from one of the 2$^\prime\times$2$^\prime$ boxes of the head region while right panel shows the spectra from the entire head region shown in figure \ref{image1} left.  The best-fit curves are shown with solid lines and the lower panels show the residuals.}
\label{spec_head} 
\end{figure}

  We found that all the spectra in the head region were remarkably similar to each other.  The spectra from one of the 2$^\prime\times$2$^\prime$ boxes in the head region and the spectra from the entire head region indicated in figure \ref{image1} are shown in figure \ref{spec_head} left and right, respectively.  We can see a prominent Si Ly$_\alpha$ line in the spectra which is consistent with those from ASCA (TMA) and Chandra (MTAM).  The high efficiency and the high energy resolution of the EPIC instrument clearly show the emission lines from O {\scshape VII} Ly$_\alpha$ triplets at 0.57\,keV and O {\scshape VIII} Ly$_\alpha$ at 0.65\,keV for the first time.  Moreover, we can see emission like structures below 0.5\,keV that corresponds to C {\scshape VI} Ly$_\alpha$ at 0.369\,keV and C {\scshape VI} Ly$_\beta$ at 0.436\,keV in the MOS-1 spectrum.  However, they cannot be seen in the MOS-2 and PN spectra (see, figure \ref{spec_head} right).  

  We fit all the spectra by an absorbed non-equilibrium ionization model (Wabs and VNEI model (NEI version 2.0) in XSPEC v\,11.3.1; Morrison \& McCammon 1983; Hamilton et al.\ 1983; Borkowski et al.\ 1994, 2001; Liedahl et al.\ 1995) with a single electron temperature, $kT_\mathrm{e}$, and a single ionization parameter, $\tau$, where $\tau$ is the number density of electron ($n_\mathrm{e}$) times the elapsed time after the shock heating.  Our free parameters are the $kT_\mathrm{e}$, $\tau$, the column density, $N_\mathrm{H}$, the emission measure (hereafter EM, EM = $\int n_\mathrm{e}n_\mathrm{i} dl$, where $n_\mathrm{i}$ is the number densities of ions, and $dl$ is the plasma depth); and abundances of C, N, O, Ne, Mg, Si, Fe.  All the abundances listed above were allowed to vary separately from each other.  We set the abundances of other elements to the solar values (Anders \& Grevesse 1989).  All the spectra are well represented by our model.  All the reduced $\chi^2$ are about unity.  We found that the best-fit parameters from all the spectra in the head region were identical to each other from the statistical point of view.  Table 1 shows the best-fit parameters of the spectra from the entire head region.  We re-fit all the spectra in the head region by the VNEI model whose parameters of abundances fixed to the values obtained for the entire head region, finding acceptable fits for all the regions ($\chi^2$/d.o.f. are 0.9$\sim$1.2).  This fact confirms the uniformity of the elemental abundances in the head region.  The temperature shown in table 1 is about 0.5\,keV which is consistent with the result by MTAM.  The values of $\tau$ measured indicate that the plasma is far from the ionization equilibrium that is consistent with the result obtained by Chandra (MTAM, 11.1$\pm$0.2).  The abundance of Si is about 3 times solar value, whereas that of O is about 0.3 times solar value, which is also consistent with the result from MTAM.  The absolute abundance predicted by the spectral fits depends on the model employed, and is strongly correlated with the other parameters.  The absolute values depend on the model while the abundance ratio between Si and O, 10$\pm 1$ is relatively reliable.

\begin{figure}
\includegraphics[angle=0,scale=.3]{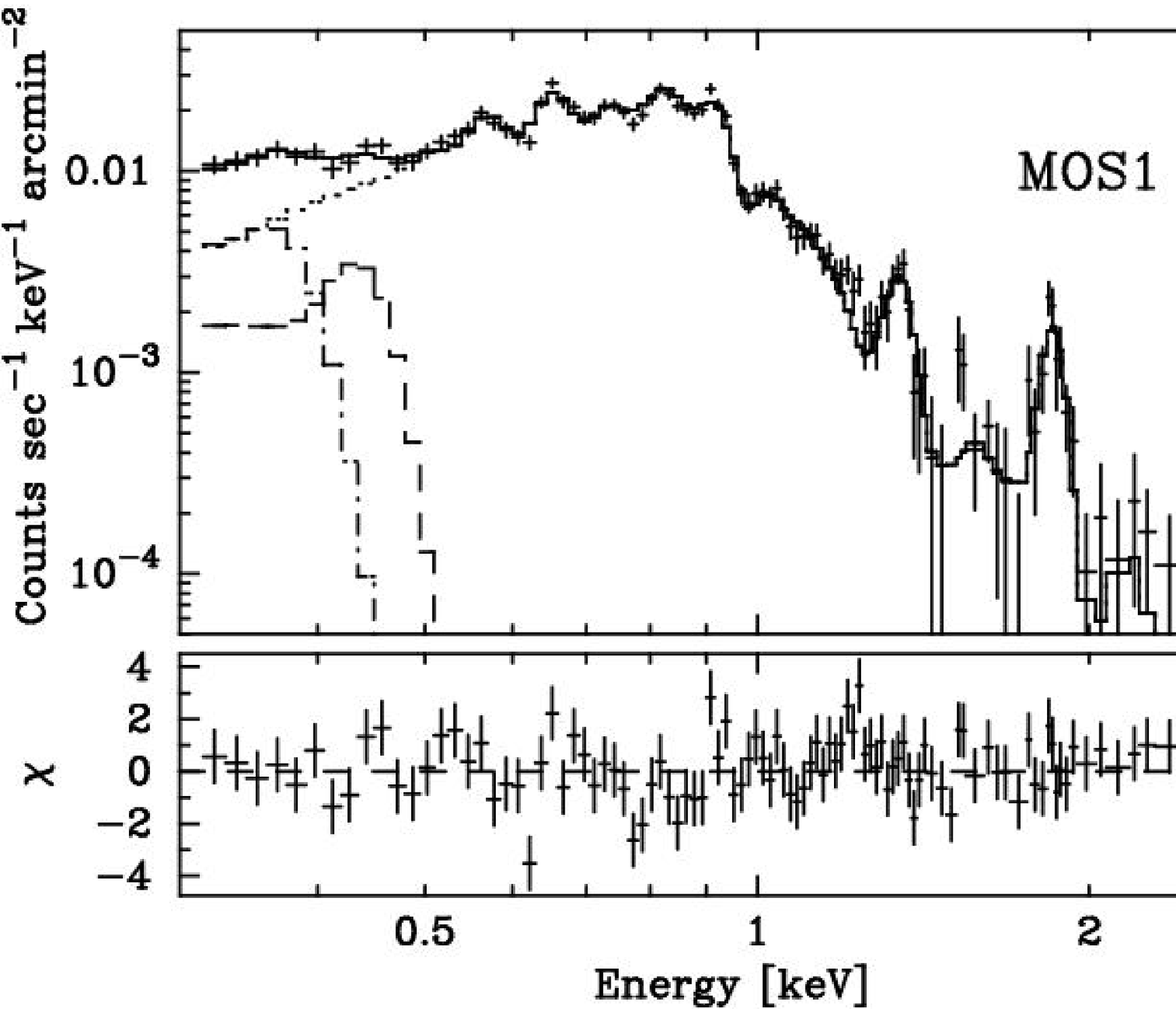}
\includegraphics[angle=0,scale=.3]{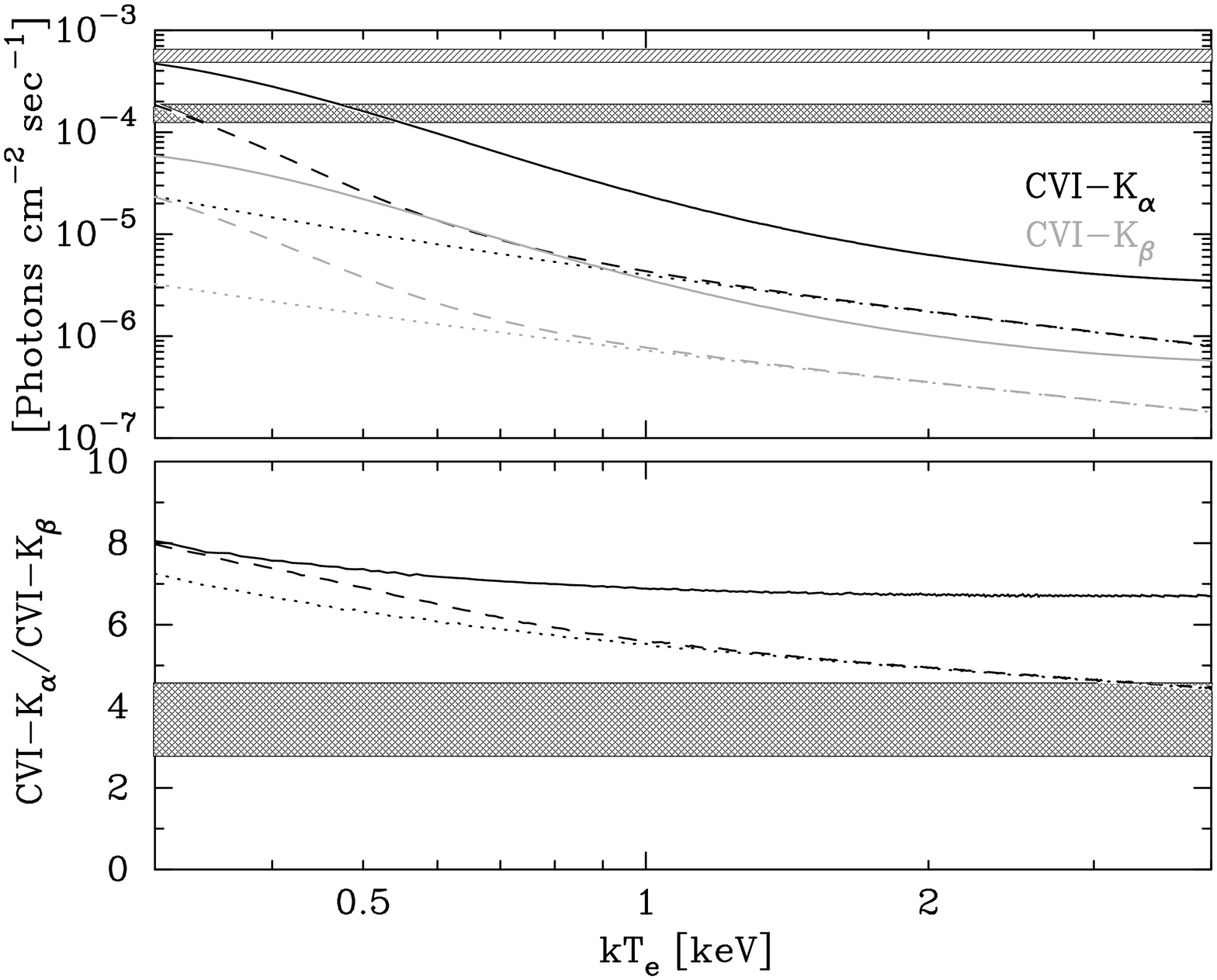}
\caption{Left: X-ray spectrum of MOS-1 from figure \ref{spec_head} right.  The solid line represents the best-fit model, while the broken lines represent the individual component.  The lower panel shows the residuals.  Right (top): The predicted fluxes of C {\scshape VI} Ly$_\alpha$ and the C {\scshape VI} Ly$_\beta$ as a function electron temperature and ionization timescale (solid, dashed and dotted correspond to 10.75, 11 and 12 of log ($\tau$), respectively).  The fluxes measured are indicated as the upper shaded region (for C {\scshape VI} Ly$_\alpha$) and the lower shaded region (for C {\scshape VI} Ly$_\beta$).  The bottom panel shows the ratio of C {\scshape VI} Ly$_\alpha$ to C {\scshape VI} Ly$_\beta$.}
\label{Cratio} 
\end{figure}

  It is doubtful that the small peaks at $\sim$0.36\,keV and $\sim$0.44\,keV really come from the emission lines from C {\scshape VI} Ly$_\alpha$ and C {\scshape VI} Ly$_\beta$, respectively since they can be seen only in the MOS-1 spectrum shown in figure \ref{spec_head} left.  Therefore, the fact that the abundance of C obtained by the VNEI model is significantly higher than the solar value (see, table 1) is also doubtful.  In order to reveal whether the two peaks seen in the MOS-1 spectrum really come from C or not, we compare the flux ratio of C {\scshape VI} Ly$_\alpha$ / C {\scshape VI} Ly$_\beta$ obtained by MOS-1 spectrum with expected values at various $kT_\mathrm{e}$ and $\tau$.  We determined the fluxes of C {\scshape VI} Ly$_\alpha$ and C {\scshape VI} Ly$_\beta$ using a combination model; a VNEI model with the abundance of C fixed to zero and two Gaussians (one for C {\scshape VI} Ly$_\alpha$ the other for C {\scshape VI} Ly$_\beta$).  We show the best-fit curve and the separated three components in figure \ref{Cratio} left.  We obtained the line flux for C {\scshape VI} Ly$_\alpha$ and C {\scshape VI} Ly$_\beta$ to be 5.5$^{+0.9}_{-0.8}$ $\times 10^{-4}$\,photons\,sec$^{-1}$\,cm$^{-2}$ and 1.6$^{+0.4}_{-0.3}$ $\times 10^{-4}$\,photons\,sec$^{-1}$\,cm$^{-2}$, respectively.  Therefore, the flux ratio of C {\scshape VI} Ly$_\alpha$ to C {\scshape VI} Ly$_\beta$ is estimated to be 3.6 $^{+1.0}_{-0.9}$.  We compare this value to that obtained by the NEI plasma model by using "{\itshape neiline}" software.  The "{\itshape neiline}" software calculates ($n_\mathrm{Z}$/$n_\mathrm{H}$)$_\odot$($n_z(\tau)$/$n_\mathrm{Z}$)$\epsilon_{i}(T)$ at constant temperature and $\tau$ where the code evolves the plasma using the Raymond \& Smith (1977; update by Brickhouse et al. 1993) plasma code, where $n_\mathrm{Z}$/$n_\mathrm{H}$ is the abundance of the element in question relative to hydrogen, $n_z(\tau)$/$n_\mathrm{Z}$ is the ionization fraction of the ionization species responsible for the line, $\epsilon_{i}(T)$, a function of temperature, is the intrinsic emissivity of the line.  We show the expected fluxes of C {\scshape VI} Ly$_\alpha$ and C{\scshape VI} Ly$_\beta$ from the plasma with solar abundance as a function of electron temperature in figure \ref{Cratio} right (top).  We calculated the emissivities for three cases; they are log($\tau$) to be 10.75, 11 and 12 where 10.75 is the best fit value in figure \ref{spec_head} right.  The emissivities measured are indicated by the shaded areas; upper for C {\scshape VI} Ly$_\alpha$) and lower for C {\scshape VI} Ly$_\beta$.  We confirmed the over abundance of C at ($kT_\mathrm{e}$, log($\tau$))=(0.52\,keV, 10.75) which were the best-fit values obtained by the VNEI model.  Figure \ref{Cratio} right (bottom) shows the ratio of C {\scshape VI} Ly$_\alpha$ to C {\scshape VI} Ly$_\beta$.  The ratio measured is plotted as the meshed region in figure \ref{Cratio} right (lower).  The flux ratio measured can be achieved only above $\sim$ 3\,keV of electron temperature whatever the ionization parameter is.  The $kT_\mathrm{e}$ is far from the value obtained by the VNEI model.  Since there is no indication of another high temperature component in the spectrum, it is natural to consider that the peaks at $0.36$\,keV and $0.44$\,keV do not come from C {\scshape VI} Ly$_\alpha$ and C {\scshape VI} Ly$_\beta$.  In conclusion, the abundance of C determined by the VNEI model is highly uncertain considering the ratio of flux.

  The difference in spectra between the MOS-1 and the MOS-2 might come from the difference of the optical blocking filters employed; MOS-1 was operated using a thin filter while MOS-2 was operated using a medium filter.  The MOS-1 spectrum is probably suffering from pile-up effects that produced spurious peaks at $0.36$\,keV and $0.44$\,keV.

\subsubsection{Tail Region}

\begin{figure}
\includegraphics[angle=0,scale=.3]{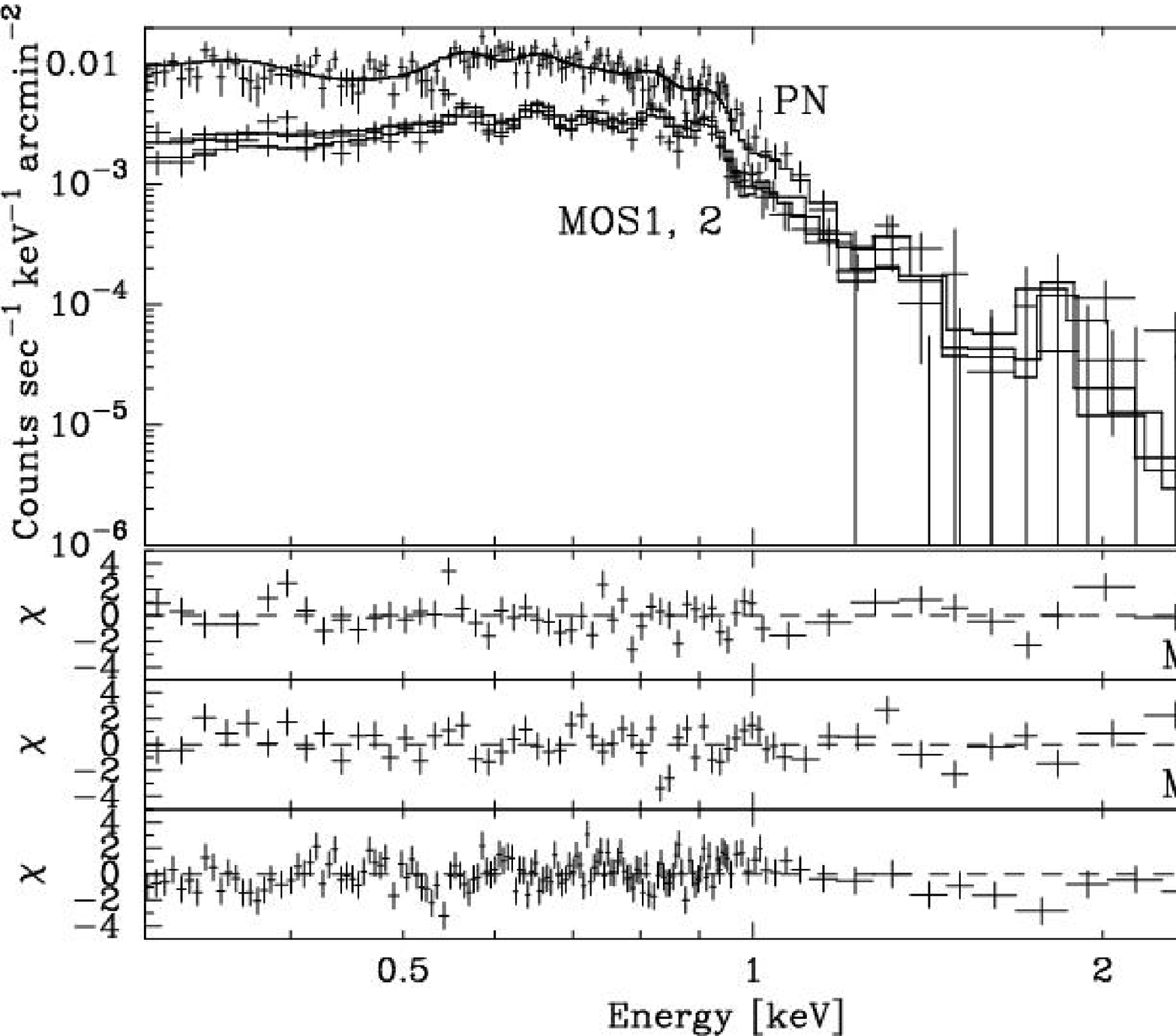}
\includegraphics[angle=0,scale=.3]{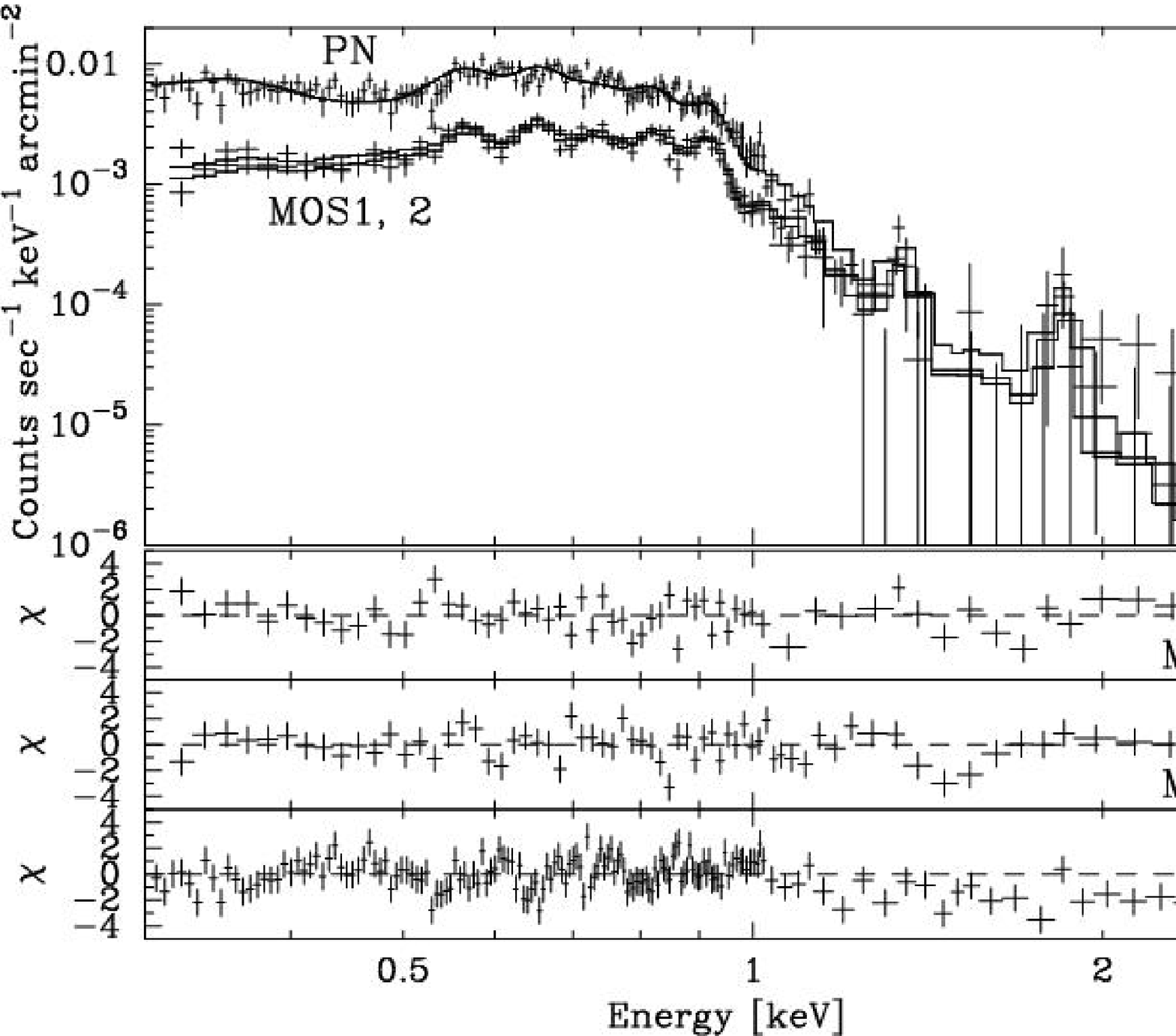}
\caption{X-ray spectra extracted from the tail region.  Left panel shows an example spectrum from one of the 4$^\prime\times$10$^\prime$ boxes of the tail region while right panel shows the spectra from the entire tail region shown in figure \ref{image1} left.  The best-fit curves are shown with solid lines and the lower panels show the residuals.}
\label{spec_tail} 
\end{figure}

  We applied an absorbed VNEI model to each spectrum and obtained relatively good fits (the reduced $\chi^{2}$ ranged from 1 to 1.3, depending on degrees of freedom from 315 to 457).  We found that the $kT_\mathrm{e}$, $\tau$ and abundances obtained by each spectrum did not show significant differences with each other.  The spectra from one of the $4^\prime\times10^\prime$ boxes in the tail region and the spectra from the entire tail region indicated in figure \ref{image1} are shown in figure \ref{spec_tail} left and right, respectively.  The spectra from the tail region are slightly different from those of the head region.  The apparent difference of the spectra between the head region and the tail region are the flux ratio of O {\scshape VII} Ly$_\alpha$ and O {\scshape VIII} Ly$_\alpha$.  The best-fit parameters for the spectrum from the entire tail region are shown in the table 1 (Tail region).  We were able to determine the abundance of Si for the tail region to be higher than solar value.  The abundance ratio of Si to O, 11 $^{+5}_{-4}$, is in good agreement with that of the head region whereas the $kT_\mathrm{e}$ is significantly lower than that of the head region.

\subsubsection{Temperature, Density and Pressure Distribution in shrapnel A}

  The electron pressure, $p_{\mathrm e}$ in the head region is estimated to be about 10 times higher than that in the tail region (AET and MTAM).  We found that $kT_{\mathrm e}$ in the tail regions are significantly lower than those of head regions.  What is the distribution of the $kT_{\mathrm e}$, $n_\mathrm{e}$ and $p_\mathrm{e}$ in shrapnel A?  In this section, we investigate the variation of the $kT_{\mathrm e}$, $n_\mathrm{e}$ and $p_\mathrm{e}$ in shrapnel A.  

\begin{figure}
\includegraphics[angle=0,scale=.4]{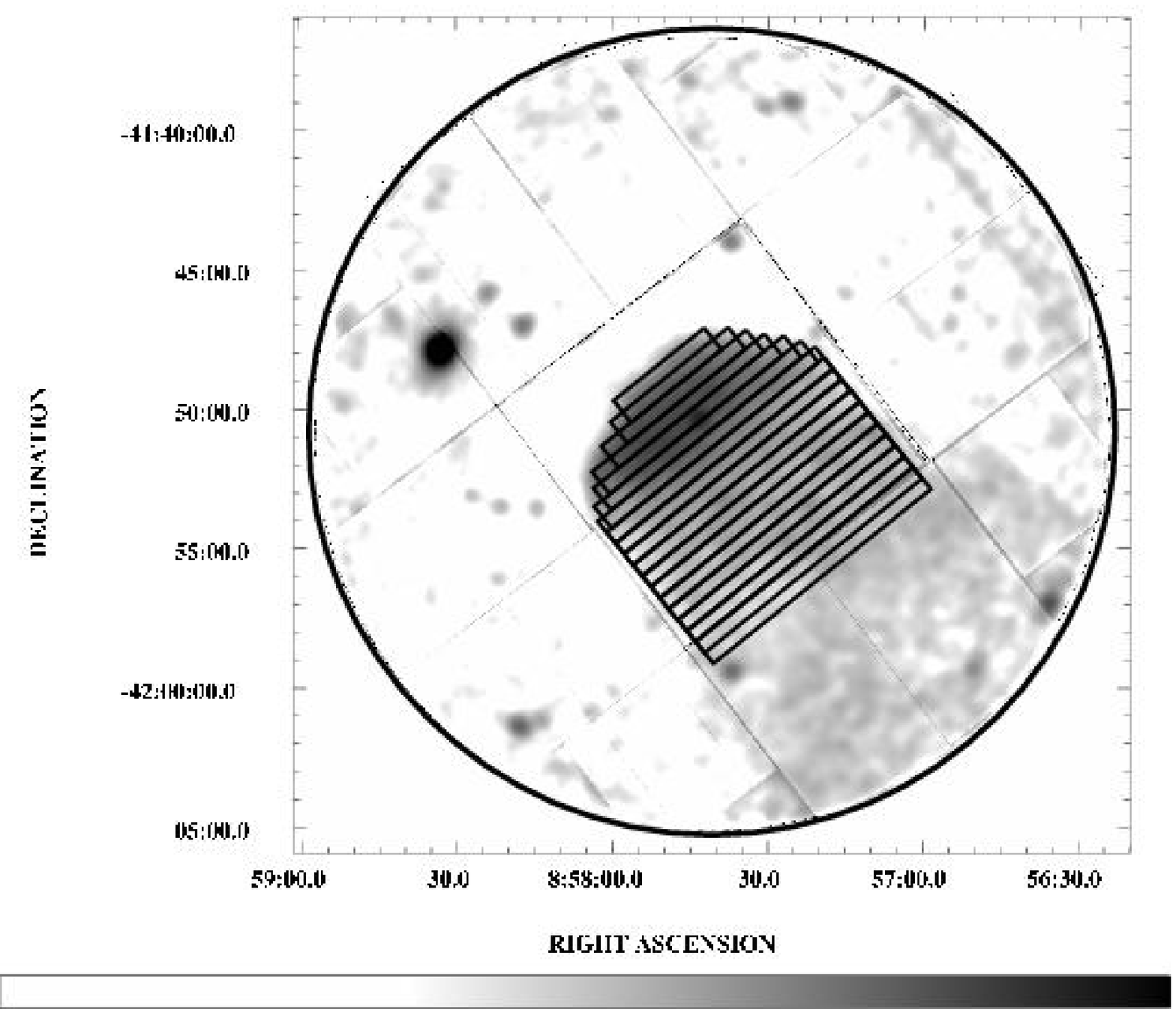}
\includegraphics[angle=0,scale=.3]{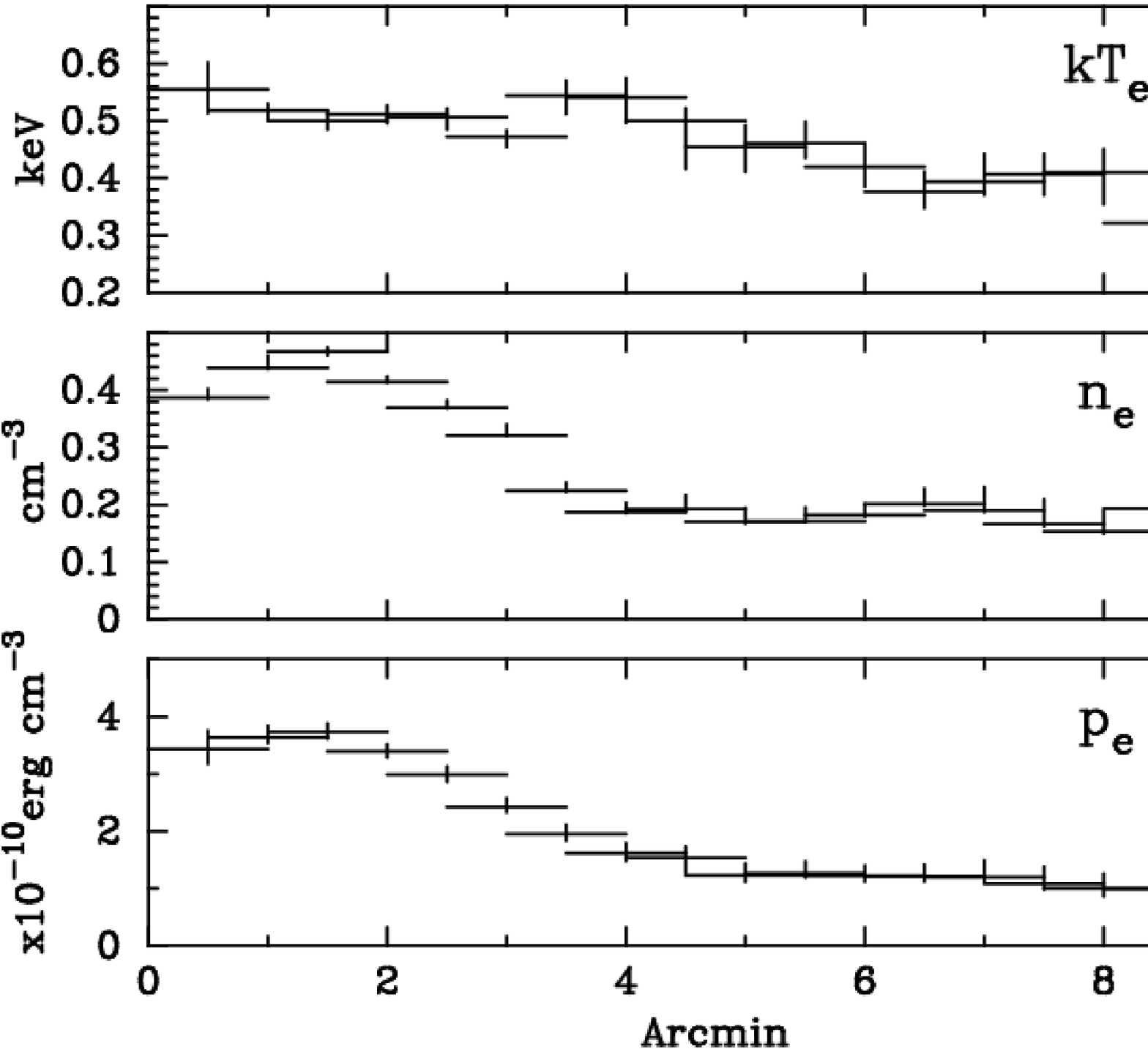}
\caption{Left: Same as figure \ref{image1} but with an overlaid the regions where we extracted spectra in order to investigate the variation of $kT_{\mathrm e}$, $n_{\mathrm e}$ and $p_{\mathrm e}$ in shrapnel A.  Right: The variation of $kT_{\mathrm e}$, $n_{\mathrm e}$ and $p_{\mathrm e}$ in the regions shown in figure \ref{image2} left.}
\label{image2} 
\end{figure}

  In order to investigate the variation, we divided shrapnel A into small rectangles shown in figure \ref{image2} left.  The width of each rectangle is selected to cover the X-ray bright region.  All the spectra are well represented by the same VNEI model as before.  We must assume the plasma depth for each region so as to estimate $n_\mathrm{e}$ and $p_\mathrm{e}$.  We assume that the shape of the shrapnel is a conical structure which is axial symmetric along the direction to the center of the Vela SNR.  It is reasonable to consider that the longer side of rectangles where we extracted spectra corresponds to the plasma depth.  Then, we can roughly estimate the plasma depth for each regions.  Assuming that the electron density is equal to that of hydrogen density, we calculated the electron density from EM and the plasma depth.  Then, we calculated $p_\mathrm{e}$ from the equation of state using $kT_{\mathrm e}$ and $n_{\mathrm e}$ obtained.  Therefore, $p_\mathrm{e}$ depends on both $kT_{\mathrm e}$ and $n_{\mathrm e}$ while $kT_{\mathrm e}$ and $n_{\mathrm e}$ are derived independently from each other.  Figure \ref{image2} right shows the variation of $kT_{\mathrm e}$, $n_{\mathrm e}$ and $p_\mathrm{e}$ for those regions.  The values of $kT_\mathrm{e}$, $n_\mathrm{e}$ and $p_\mathrm{e}$ gradually decrease from head region to tail region.  There is another estimation of the plasma depth for the tail region.  Based on the figures from Jones et al. (1994) and Anderson et al. (1994), almost all the gas in the tail region is compressed in a shell region after the shock front.  The depth of the shell region is lower than one quarter of the radius of the conical structure of the shrapnel so that the plasma depth for the tail region approximately reduces to lower than one quarter of that estimated above.  If it is the case, $n_\mathrm{e}$ and $p_\mathrm{e}$ for the tail region increase by at least factor 2.

\section{Discussion and Conclusion}

  We performed spatially resolved spectral analysis of Vela shrapnel A with the XMM-Newton satellite.  The X-ray image clearly reveals that shrapnel A consists of a bright knot (head region) and a faint trailing wake (tail region).  We found that the spectra are similar to each other at any portion in shrapnel A.  They are represented well by a single-temperature NEI model.  We confirmed the abundance of Si was a few times higher than that of solar value while other metal abundances are solar or subsolar values; O$\sim$0.3, Ne$\sim$0.9, Mg$\sim$0.8, Si$\sim$3, Fe$\sim$0.8.  The absolute abundances depend on the model employed while the relative abundances between heavy elements are relatively robust.  Therefore, it is reasonable to consider that shrapnel A is extremely rich in Si.  We estimate masses of each element relative to O.  They are $\sim$0.5, $\sim$0.2, $\sim$0.7, $\sim$0.5, for Ne, Mg, Si and Fe, respectively.  Thielemann et al. (1996) calculated the isotopic composition of the ejecta of core-collapse SNe from 13, 15, 20 and 25 $\mathrm{M_{\odot}}$ stars.  We compared the mass ratio relative to O with those models.  We can easily find that there is no such layer satisfying the ratio obtained.  This indicates that shrapnel A is strongly contaminated by the swept-up ISM.  TMA presumed that Vela SN occurred in a bubble of hot tenuous gas which was considered to be an old SNR.  Therefore, the abundance of the swept-up ISM might be different from solar values so that it is difficult to estimate the masses of metals from swept-up ISM  in shrapnel A.  O and Si are the most massive elements in shrapnel A so that they might reflect the initial composition.  When we employ only the relative mass of O to Si for comparison with theoretical models, shrapnel A turns out to come from the layer inside the progenitor star whose mass radius is 1.55$\sim$2 $\mathrm{M_{\odot}}$ (depending on the mass of the progenitor star).

  The VNEI model indicates an extreme over abundance of C in shrapnel A.  In our data, we found that the plasma with temperature beyond 3\,keV could explain the measured flux ratio of C {\scshape VI} Ly$_\alpha$ to C {\scshape VI} Ly$_\beta$.  There is no other indication of existence of such a high temperature plasma.  This fact suggests that the abundance of C determined by the VNEI model is highly uncertain, probably due to the uncertainty of the calibration below 0.5\,keV.  A new X-ray astronomy satellite, "Suzaku", was launched in July, 2005.  The CCD camera, XIS, on board Suzaku has high performance below 0.5\,keV that will clearly detect C emission lines if they are there.

  We found that the electron temperature, the electron density and the electron pressure gradually decreases toward the shell of the Vela SNR.  If the hot plasma is generated by a simple blast wave, the temperature increases toward the center of the explosion.  Therefore, the variation of $kT_\mathrm{e}$ supports the idea that shrapnel A originates not from the blast wave but from the explosion ejecta.  A similar structure in temperature is seen in shrapnel D (Katsuda \& Tsunemi. 2005).  Therefore, the temperature variation appears to be a common property of the explosion ejecta.  The electron pressure in the head region was estimated to be about 10 times higher than that of the tail region from the observation of ROSAT (AET) and ASCA (MTAM).  However, we estimated the electron pressure to be $\sim$4$\times$ 10$^{-10}$ erg\,cm$^{-3}$ at the top of shrapnel A and decreases down to $\sim$1$\times$ 10$^{-10}$ erg\,cm$^{-3}$ i.e. the electron pressure decreases only by a factor of 4.  If we employ a thick hollow conical structure in the tail region (Jones et al. 1994, Anderson et al. 1994), the decreases are reasonably reduced to lower than a factor of 2.  

  We assume that the head region of shrapnel A has a conical structure with a diameter 9$^\prime$ and a height 4$^\prime$ and the tail region has a cylinder with a diameter 10$^\prime$ and a height 11$^\prime$.  We estimate the X-ray emitting volume for head region and tail region to be $\sim$1$\times$10$^{54}$cm$^{3}\times(d/250)^{3}$ and $\sim$1$\times$10$^{55}$cm$^{3} (d/250)^{3}$, respectively, where $d$ is the distance to Vela shrapnel A in units of pc.  Using the density, 0.5 for head regions and 0.2 for tail regions, mass of shrapnel A is estimated to be $\sim$5$\times$10$^{-3}\mathrm{M_{\odot}}$, which is smaller than that of shrapnel D (0.1$\times\mathrm{M_{\odot}}$).  There are fast-moving knots in Cas A that show a pure O or O with products of O burning, S, Ar, Ca (Kirshner \& Chevalier 1977, Chevarier \& Kirshner 1978, 1979).  The optically emitting mass of the knots in Cas-A is only about $10^{-4}\mathrm{M_{\odot}}$ (Raymond 1984).  The masses of shrapnel A and D are larger than those of the optically emitting knots found in the Cas-A SNR.  The masses of shrapnels A and D were estimated on the assumption that they mainly consists of the hydrogen.  However, the shrapnels might be lacking in hydrogen instead consisting purely of metal from the progenitor star.  If it is the case, the heavy elements make an important contribution to the continuum emission, which causes a change of the inferred mass of X-ray emitting gas (e.g., Vink et al. 1996).  If we assume that shrapnel A consists of only metal, then the mass will reduce to one quarter of that obtained above. 

  The interaction of clumps with a uniform medium has been studied in two-dimensional simulations by Anderson et al. (1994) and Jones et al (1994).  They identified in the evolution of clumpy ejecta: a bow-shock phase, an instability phase and a dispersal phase.  The three evolutionary phases will occur well before the nominal ram pressure "stopping time", defined as the timescale over which the clump intercepts its own mass from the ambient medium (Anderson et al. 1994).  If shrapnel A has been keeping its shape and sweeping-up the ambient medium since the explosion, the shrapnel turns out to sweep-up the ambient medium by $\rho_0$ L S $=\rho_0$ (L/$d$) (S/$d^{\mathrm 2}) d^{\mathrm 3}$, where $\rho_0$, L/d, S/$d^{\mathrm 2}$ are the density of ambient medium, the angular distance from the center of the Vela SNR, the angular size of the shrapnel.  Using the number density of the ambient medium estimated by TMA, we can estimate the mass of the swept-up ambient medium to be $\sim$7$\times$10$^{-3}$ ($\rho_0/0.02$ cm$^{\mathrm -3}$)(5.3$^{\circ}$)($\pi \times36$ arcmin$^2$)($d$/250 pc)$^3$ $\mathrm{M_{\odot}}$.  This value exceeds the obtained mass of shrapnel A, which indicates that the shrapnel should be in the dispersal phase.  Nonetheless, the shrapnel clearly shows a bright X-ray knot feature that is defined as the head region here, which implies that the shrapnel has not reached the dispersal phase yet.  Therefore, the view studied by Wang et al. (2002) might be the case, that is shrapnel A is comoving with the surrounding ejecta for a long time and passed over a blast wave.

\acknowledgments

The authors would express their special thanks to Prof. R. Smith for kindly giving us the software, "{\itshape neiline}".  This work is partly supported by a Grant-in-Aid for Scientific Research by the Ministry of Education, Culture, Sports, Science and Technology (16002004).  This study is also carried out as part of the 21st Century COE Program, \lq{\it Towards a new basic science: depth and synthesis}\rq.

\clearpage

\begin{deluxetable}{cccrrrrrrrcrl}
\tabletypesize{\scriptsize}

\tablecaption{Spectral-fit parameters.}
\tablewidth{0pt}
\tablehead{
\colhead{Parameter} & \colhead{Head region in figure \ref{image1}}  & \colhead{Tail region in figure \ref{image1}}
}
\startdata
   $N_{\mathrm H} [10^{20}\,\mathrm{cm}^{-2}$] \dotfill & 3.2$^{+1.4}_{-0.4}$ &1.4$^{+0.02}_{-0.01}$ \\
   $kT_\mathrm{e}$[keV] \dotfill &0.52$\pm 0.01$&0.37$\pm 0.01$  \\
   C  \dotfill  & 2.5$\pm 0.2$ & 2.6$^{+0.5}_{-0.6}$ \\
   N  \dotfill  & 0.55$^{+0.06}_{-0.08}$ & 0.5$^{+0.2}_{-0.1}$ \\
   O  \dotfill  & 0.34$\pm 0.01$ & 0.4$^{+0.01}_{-0.02}$ \\
   Ne \dotfill  & 1.07$\pm 0.04$ & 1.28$^{+0.09}_{-0.07}$  \\
   Mg \dotfill  & 0.87$\pm 0.08$ & 0.96$^{+0.4}_{-0.3}$  \\
   Si \dotfill  & 3.3$\pm 0.3$ & 3.7$^{+0.7}_{-1.0}$  \\
   Fe \dotfill  & 0.96$\pm 0.03$ & 1.10$^{+0.08}_{-0.07}$ \\
   log($\tau$) $[\mathrm{s\,cm^{-3}}]$ \dotfill & 10.75$\pm 0.02$ & 10.97$^{+0.04}_{-0.03}$  \\ 
   EM$[\mathrm{cm^{-5}}]^{\mathrm a}$\dotfill&(2.35$^{+0.1}_{-0.03}$) $\times10^{17}$ & $(0.53^{+0.04}_{-0.02})\times10^{17}$\\ 
\hline
   $\chi^2$/d.o.f. \dotfill & 705/465 & 698/544 \\
\enddata

\tablecomments{Other elements are fixed to those of solar values.  The values of abundances are multiples of solar value.  The errors are in the range $\Delta\,\chi^2\,<\,2.7$ on one parameter.}
\tablenotetext{a}{EM denotes the emission measure $\int n_\mathrm{e} n_\mathrm{i} dl$.}
\end{deluxetable}

\begin{thebibliography}{}
\bibitem[Anders(1989)]{Anders1989} 
	Anders, E., \& Grevesse, N. 1989, Geochim. Cosmochim. Acta, 53, 197
\bibitem[Anderson(1994)]{Anderson1994}
	Anderson, M. C., Jones, T. W., Rudnick, L., Tregillis, I. L., \& Kang, H. 1994, ApJ, 421, L31
\bibitem[Aschenbach(1995)]{Aschenbach1995} 
	Aschenbach, B., Egger, R., \& Trumper, J. 1995, Nature, 373, 587 
\bibitem[Baskill(2005)]{Baskill2005}
	Baskill, D. S., Wheatly, P. J., \& Osborne, J. P. 2005, MNRAS, 357, 626
\bibitem[Borkowski(2001)]{Borkowski2001} 
	Borkowski, K. J., Lyerly W. J., \& Reynolds, S. P. 2001, ApJ, 548, 820
\bibitem[Borkowski(1994)]{Borkowski1994} 
	Borkowski, K. J., Sarazin, C. L., \& Blondin, J. M. 1994, ApJ, 429, 710 
\bibitem[Brickhouse(1993)]{Brickhouse1993} 
	Brickhouse, N. S., Raymond, J. C., \& Smith, B. W. 1993, BAAS, 25, 864
\bibitem[Cha(1999)]{Cha1999} 
	Cha, A. N., Sembach, K. R., \& Danks, A. C. 1999, ApJ, 515, L25
\bibitem[Chevalier(1978)]{Chevalier1978} 
	Chevalier, R. A., \& Kirsher, R. P. 1978, ApJ, 219, 931
\bibitem[Chevalier(1979)]{Chevalier1979} 
	Chevalier, R. A., \& Kirsher, R. P. 1979, ApJ, 233, 154
\bibitem[Decourchelle(2001)]{Decourchelle2001} 
	Decourchelle, A., et al. 2001, A\&A, 365, L218
\bibitem[Fesen(2002)]{Fesen2002} 
	Fesen, R. A. 2001, ApJS, 133, 161
\bibitem[Gansicke(1999)]{Gansicke1999}
	Gansicke, B. T., \& Koester, D. 1999, A\&A, 346, 151
\bibitem[Hamilton(1983)]{Hamilton1983} 
	Hamilton, A. J. S., Sarazin, C. L., \& Chevalier, R. A. 1983, ApJS, 51, 115
\bibitem[Jones(1994)]{Jones1994}
	Jones, T. W., Kang, H., \& Tregillis, I. L. 1994, ApJ, 432, 194
\bibitem[Katsuda(2005)]{Katsuda2005}
	Katsuda, S., \& Tsunemi, H. 2005, PASJ, 57, 621
\bibitem[Kirshner(1977)]{Kirshner1977} 
	Kirshner, R. P., Chevalier, R. A. 1977, ApJ, 218, 142
\bibitem[Liedahl(1995)]{Liedahl1995} 
	Liedahl, D. A., Osterheld, A. L., \& Goldstein, W. H. 1995, ApJ, 438, L115
\bibitem[Mewe(1986)]{Mewe1986}
	Mewe, R., Lemen, J. R., van den Oord G. H. J., 1986, A\&AS, 65, 511
\bibitem[Miyata(2001)]{Miyata2001} 
	Miyata, E., Tsunemi, H., Aschenbach, B., \& Mori, K. 2001, ApJ, 559, L45
\bibitem[Morrison(1983)]{Morrison1983} 
	Morrison, R., \& McCammon, D. 1983, ApJ, 270, 119
\bibitem[Raymond(1984)]{Raymond1984}
	Raymond, J. C. 1984, ARA\&A, 22, 75
\bibitem[Raymond(1977)]{Raymond1977} 
	Raymond, J. C., \& Smith, B. W., 1977, ApJS, 35, 419
\bibitem[Taylor(1993)]{Taylor1993} 
	Taylor, J. H., Manchester, R. N. \& Lyne, A, G. 1993, ApJS, 88, 529
\bibitem[Thielemann(1996)]{Thielemann} 
	Thielemann, F.-K., Nomoto, K., \& Hashimoto, M., 1996, ApJ, 460, 408
\bibitem[Tsunemi(1999)]{Tsunemi1999}
	Tsunemi, H., Miyata, E., \& Aschenbach, B. 1999, PASJ, 51, 711
\bibitem[Verbunt(1997)]{Verbunt1997}
	Verbunt, F., Bunk, W. H., Ritter, H., Pfeffermann, E., 1997, A\&A, 327, 602
\bibitem[Vink(1996)]{Vink1996}
	Vink, J., Kaastra, J. S., \& Bleeker, J. A. M. 1996, A\&A, 307, 41
\bibitem[Wang(2002)]{Wang2002}
	Wang, C. Y., \& Chevalier, R. A. 2002, ApJ, 574, 155
\bibitem[Warner(1987)]{Warner1987}
	Warner, B., 1987, MNRAS, 227, 23
\end{thebibliography}
\end{document}